\documentclass{aastex}
\usepackage{spr-astr-addons}
\usepackage{graphicx}
\usepackage{txfonts}
\usepackage{xcolor}
\usepackage{tikz}
\usepackage{rotating}
\usepackage{lscape}
\usepackage{adjustbox}
\usepackage[version=4]{mhchem}
\usepackage[normalem]{ulem}
\usepackage{float}
\usepackage{natbib}
\pdfoutput=1 
\usepackage{amsmath,amstext,amssymb}
\usepackage[T1]{fontenc}
\usepackage[colorlinks=true, allcolors=blue]{hyperref}

\RequirePackage{color}
\begin{document}
	
\title{Detection of complex nitrogen-bearing molecule ethyl cyanide towards the hot molecular core G10.47+0.03}
	\shorttitle{Ethyl cyanide in G10.47+0.03}
	\shortauthors{Manna \& Pal}

	\author{Arijit Manna\altaffilmark{1}} \and \author{Sabyasachi Pal\altaffilmark{1}}
	\email{arijitmanna@mcconline.org.in}
	
	\altaffiltext{1}{Department of Physics and Astronomy, Midnapore City College, Paschim
		Medinipur, West Bengal, India 721129 \\email: {arijitmanna@mcconline.org.in}}

	\begin{abstract}
The studies of the complex organic molecular lines towards the hot molecular cores at millimeter and submillimeter wavelengths provide instructive knowledge about the chemical complexity in the interstellar medium (ISM). We present the detection of the rotational emission lines of the complex nitrogen-bearing molecule ethyl cyanide (C$_{2}$H$_{5}$CN) towards the chemically rich hot molecular core G10.47+0.03 using the Atacama Large Millimeter/Submillimeter Array (ALMA) band 4 observations. The estimated column density of C$_{2}$H$_{5}$CN towards the G10.47+0.03 is (7.7$\pm$0.5)$\times$10$^{16}$ cm$^{-2}$ with the high rotational temperature of 352.9$\pm$66.8 K. The estimated fractional abundance of C$_{2}$H$_{5}$CN with respect to H$_{2}$ towards the G10.47+0.03 is 5.70$\times$10$^{-9}$. We observe that the estimated fractional abundance of C$_{2}$H$_{5}$CN is similar to the existing three-phase warm-up chemical modelling abundance of \ce{C2H5CN}. We also discuss the possible formation mechanism of C$_{2}$H$_{5}$CN towards the hot molecular cores, and we claim the barrierless and exothermic radical-radical reaction between \ce{CH2} and \ce{CH2CN} is responsible for the production of low abundant of \ce{C2H5CN} ($\sim$10$^{-9}$) in the grain surface of G10.47+0.03.
	
\end{abstract}


\section{Introduction}
\label{sec:intro} 
In the ISM, or circumstellar shells, more than 270 interstellar complex and prebiotic molecules are identified at millimeter and submillimeter wavelengths\footnote{\href{https://cdms.astro.uni-koeln.de/classic/molecules}{https://cdms.astro.uni-koeln.de/classic/molecules}}. The identification of the molecular lines from the ISM is a key step in the understanding of the chemical evolution from simple molecular species to molecules of biological relevance \citep{her09}. The hot molecular cores are known as the early processes of the high-mass star-formation region, which play a key role in increasing the chemical complexity in the ISM \citep{shi21}. At this stage, the complex organic molecules are ejected from the icy surfaces of dust grains \citep{her09}. The hot molecular cores are characterised by their warm temperature ($\gtrsim$100 K), small source size ($\leq$0.1 pc), and high gas density ($\gtrsim$10$^{6}$ cm$^{-3}$) \citep{van98, kur00}. The time scale of the hot molecular cores lies between $\sim$10$^{5}$ years and $\sim$10$^{6}$ years \citep{van98, gar06, gar13}. The hot molecular cores contain the high-velocity water (H$_{2}$O) maser emission, which is located near the ultra-compact (UC) H II regions \citep{meh04, man22a}.  The hot core phase is identified by chemically rich molecular emission spectra with many complex organic molecules, including methyl cyanide (CH$_{3}$CN) and methanol (CH$_{3}$OH) \citep{al17}. These molecules may be created in the hot molecular cores at a high temperature ($\gtrsim$150 K) via endothermic chemical reactions \citep{al17}. Using high spectral and spatial resolution data like ALMA, VLA, etc., we can identify the different complex organic molecules and the spatial distribution of those complex molecules in the hot molecular cores. The identification of the disk candidates in the hot molecular cores is very rare, which suggests a correlation between disks and hot molecular core chemistry \citep{al17}. The study of the molecular lines in the disk candidate hot molecular cores can help us understand the high-mass star-formation process and chemical evaluation on the small physical scales ($\leq$0.05--0.1 pc) \citep{al17}.
	
\begin{table*}{}
	\centering
	\caption{Observation summary of G10.47+0.03.}
	\begin{adjustbox}{width=1.0\textwidth}
		\begin{tabular}{cccccccccccc}
			\hline
			Observation date &	On-source time&Frequency range&Spectral resolution &Sensitivity (10 km s$^{-1}$)& Angular resolution\\	
			(yyyy-dd-mm)     &	(hh:mm)       &(GHz)            & (kHz)                 &(mJy beam$^{-1}$)&($^{\prime\prime}$)\\
			\hline		
			2017-03-05       &	01:53.40      &129.50--130.44&1128.91               &	0.61&1.67$^{\prime\prime}$ (14362 au) 	\\
			--              &   --            &130.09--130.56 &~488.28                            &0.61&\\
			--               &   --           &130.54--131.01 &~488.28     &0.62&\\
			--              &   --            &130.50--131.44&1128.91   &0.62&\\

			\hline
			2017-01-28       &	00:33.06     &147.50--148.43&1128.91               &	0.50&1.52$^{\prime\prime}$ (13072 au) 	\\
			--              &   --           &147.55--148.01 &~488.28                  &0.50&\\
			--              &   --           &148.00--148.47&~488.28   &0.53&\\
			--               &   --          &148.50--149.43 &1128.91     &0.53&\\
			
			\hline
			2017-03-06       &	01:03.50      &153.00--153.93&1128.91               &	0.66&1.66$^{\prime\prime}$ (14276 au)	\\
			--              &   --            &153.36--153.83 &~488.28                            &0.66&\\
			--              &   --            &153.82--154.29 &~488.28   &0.67&\\
			--               &   --           &154.00--154.93 &1128.91      &0.67&\\
			
			\hline
			2017-03-07       &	00:28.72      &158.49--159.43 &1128.91               &	0.78&1.76$^{\prime\prime}$ (15136 au) 	\\
			--              &   --           &159.18--159.65 &~488.28                            &0.80&\\
			--              &   --           &159.49--160.43 &1128.91   &0.79&\\
			--               &   --           &159.64--160.11 &~488.28     &0.79&\\
			
			\hline
		\end{tabular}	
	\end{adjustbox}

	\label{tab:data}
\end{table*} 

The complex organic molecules that contain --C$\equiv$N functional group play a key role in prebiotic chemistry because this functional group is responsible for the formation of peptides, nucleic acids, amino acids, and nucleobases of DNA and RNA in the universe \citep{bel09}. The complex nitrogen-bearing molecule ethyl cyanide (C$_{2}$H$_{5}$CN) is also known as propanenitrile, and this is a well-known interstellar molecule which is found in the hot molecular cores and the massive star-formation regions \citep{meh04}. The dipole moments of the a- and b-conformers of the asymmetric top molecule C$_{2}$H$_{5}$CN are $\mu_{a}$ = 3.85 D and $\mu_{b}$ = 1.23 D, respectively \citep{he74}. The high dipole moment of \ce{C2H5CN} indicates that it exhibits a high intensity and dense rotational spectrum. The \ce{C2H5CN} molecules form on the dust grains of high gas density, the warm inner region of the hot molecular cores \citep{meh04}. The emission lines of C$_{2}$H$_{5}$CN were first detected towards the Orion molecular cloud (OMC-1) and Sgr B2, with estimated column densities of 1.8$\times$10$^{14}$ cm$^{-2}$ and 1.6$\times$10$^{14}$ cm$^{-2}$, respectively \citep{jon77}. Later, hundreds of the rotational emission lines of C$_{2}$H$_{5}$CN were detected towards the high-mass star-formation regions Sgr B2, Orion, and W51 \citep{mio97, li01}. The column density of C$_{2}$H$_{5}$CN towards the Sgr B2 (N) large molecule heimat source or Sgr B2 (N-LMH) hot core reaches 1$\times$10$^{17}$ cm$^{-2}$ \citep{mio97}. The higher-excited transitional lines of C$_{2}$H$_{5}$CN were also detected in the hot molecular cores Orion KL and Sgr B2 (N) \citep{dal13, bel13}. The vibrationally excited transition lines of C$_{2}$H$_{5}$CN were also detected in the Sgr B2 and in W51 e2 \citep{meh04, dem08}. Earlier, three $^{13}$C isotopologues of ethyl cyanide, like $^{13}$CH$_{3}$CH$_{2}$CN, CH$_{3}$$^{13}$CH$_{2}$CN, and CH$_{3}$CH$_{2}$$^{13}$CN were identified in the Orion hot molecular cloud in the frequency ranges of 80--40 GHz and 160--360 GHz \citep{dem07}. The evidence of the C$_{2}$H$_{5}$CN was also found in the low-mass protostars NGC 1333-IRAS 4A, NGC 1333-IRAS 2A, and IRAS 16293--2422 \citep{caz03, taq15}. Recently, evidence of the emission lines of C$_{2}$H$_{5}$CN was also found in the atmosphere of Titan using the ALMA \citep{cor14, man22b}.

\begin{table*}{}
	\centering
	\caption{Summary of the continuum image properties of G10.47+0.03.
	}
	\begin{adjustbox}{width=1.0\textwidth}
		\begin{tabular}{cccccccccccc}
			\hline
			Observation date&		Frequency&Integrated flux&Peak flux &Beam size &Position angle& RMS\\
			(yyyy-dd-mm) &			(GHz)   &(Jy)    &(Jy beam$^{-1}$)&($^{\prime\prime}$$\times$$^{\prime\prime}$)&($^{\circ}$)&(mJy beam$^{-1}$)\\
			\hline
			2017-03-05&130.234&1.36$\pm$0.01&1.08$\pm$0.08&2.38$\times$1.54&71.35&8.66\\
			&130.320&1.31$\pm$0.02&1.02$\pm$0.07&2.36$\times$1.54&72.41&8.50\\
			&130.773&1.30$\pm$0.01&1.02$\pm$0.08&2.37$\times$1.52&71.67&5.24 \\
			&130.949&1.51$\pm$0.03&1.16$\pm$0.08&2.35$\times$1.52&72.15&2.72 \\
			\hline
			2017-01-28 &147.692&1.96$\pm$0.03&1.37$\pm$0.01&2.02$\times$1.61&70.06&12.81 \\			
			&147.992&2.06$\pm$0.03&1.44$\pm$0.01&2.01$\times$1.60&69.84&14.20 \\
			&148.237&2.22$\pm$0.03&1.56$\pm$0.03&2.01$\times$1.60&69.47&~~9.74 \\
			&149.044&2.32$\pm$0.03&1.59$\pm$0.01&2.07$\times$1.60&69.73&16.23 \\
			\hline
			2017-03-06 &153.486&2.32$\pm$0.02&1.76$\pm$0.02&2.10$\times$1.38&77.92&~~4.61 \\
			&153.597&2.11$\pm$0.02&1.62$\pm$0.01&2.11$\times$1.39&77.89&~~9.08 \\
			&154.038&2.08$\pm$0.03&1.58$\pm$0.02&2.17$\times$1.39&80.86&~~9.92 \\
			&154.482&2.20$\pm$0.02&1.72$\pm$0.04&2.08$\times$1.39&78.13&~~2.86 \\
			\hline
			2017-03-07&			158.942&2.43$\pm$0.03&1.91$\pm$0.01&2.42$\times$1.44&77.48&15.59 \\
			&159.414&2.39$\pm$0.04&1.87$\pm$0.02&2.46$\times$1.42&75.54&~~7.67 \\
			&159.928&2.39$\pm$0.04&1.87$\pm$0.01&2.41$\times$1.45&78.01&~~5.15 \\
			&160.155&2.71$\pm$0.05&2.09$\pm$0.02&2.41$\times$1.44&77.74&15.59 \\
			\hline
		\end{tabular}	
	\end{adjustbox}	
	
	\label{tab:cont}
\end{table*}

The hot molecular core G10.47+0.03 is known as a UC H II region, which is located at a distance of 8.6 kpc with a luminosity of 5$\times$10$^{5}$ L$_{\odot}$ \citep{ ce10, san14}. Earlier, the rotational emission lines of complex organic molecules such as methyl isocyanate (CH$_{3}$NCO), formamide (NH$_{2}$CHO), glycolaldehyde (HOCH$_{2}$CHO), methyl isocyanate (HNCO), ethylene glycol ((CH$_{2}$OH)$_{2}$), methanol (CH$_{3}$OH), propanal (CH$_{3}$CH$_{2}$CHO), acetone (CH$_{3}$OCH$_{3}$), methyl formate (CH$_{3}$OCHO), dimethyl ether (CH$_{3}$COCH$_{3}$), and acetaldehyde (CH$_{3}$CHO) were also found towards the G10.47+0.03 \citep{rol11, gor20, mondal21}. The rotational emission lines of possible glycine (\ce{NH2CH2COOH}) precursor molecules such as methenamine ({CH$_{2}$NH}), methylamine ({CH$_{3}$NH$_{2}$}), and amino acetonitrile ({NH$_{2}$CH$_{2}$CN}) were also detected from the hot molecular core G10.47+0.03 using the Nobeyama 45 m radio telescope and ALMA band 4 \citep{suz16, ohi19, man22c}. Recently, the emission lines of the amide-type molecule cyanamide ({NH$_{2}$CN}) were detected from G10.47+0.03 using the ALMA with an estimated column density of (6.60$\pm$0.1)$\times$10$^{15}$ cm$^{-2}$ \citep{man22d}. The amide-related molecule {NH$_{2}$CN} was previously identified in the ISM as a possible precursor of urea {(NH$_{2}$CONH$_{2}$)} \citep{man22d}.

\begin{figure*}
	\centering
	\scriptsize
	\includegraphics[width=1.0\textwidth]{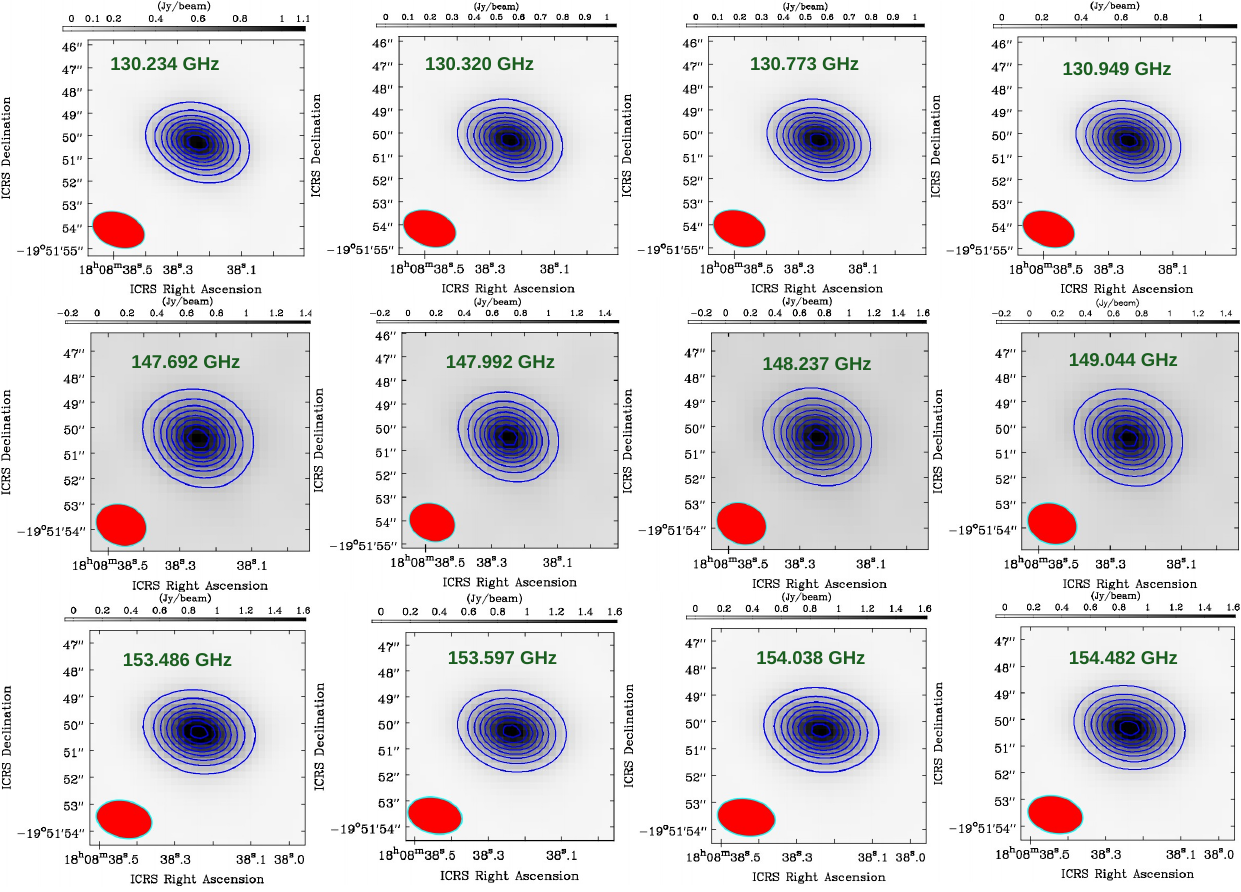}
	\includegraphics[width=1.0\textwidth]{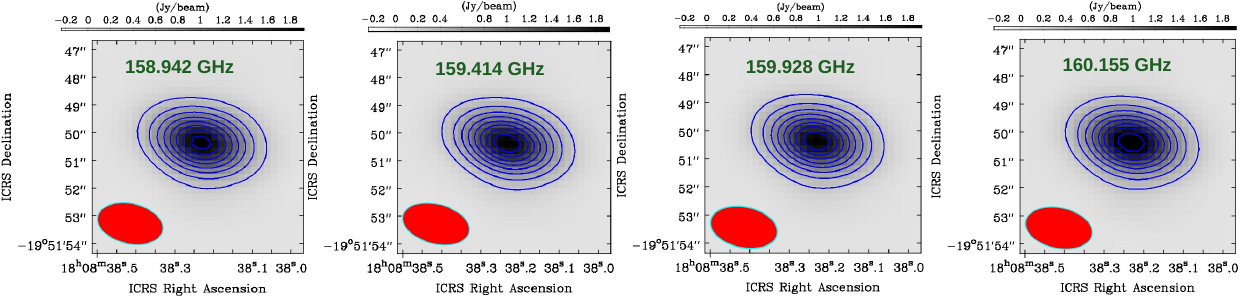}	
	\caption{Millimeter-wavelength continuum emission images of the hot molecular core G10.47+0.03 obtained with ALMA band 4 in the frequency range 130.234 GHz--160.155 GHz. The contour levels start at 3$\sigma$, where $\sigma$ is the RMS of each continuum image, and the contour levels increase by a factor of $\surd$2. The red circles indicate the synthesised beam of the continuum images. The corresponding synthesised beam size and RMS of all continuum images are presented in Table.~\ref{tab:cont}.}
	\label{fig:continuum}
\end{figure*}

In this article, we present the identification of the rotational emission lines of the complex nitrile-bearing molecule ethyl cyanide (C$_{2}$H$_{5}$CN) towards the G10.47+0.03 using the ALMA. The observations and data reduction of the raw data of G10.47+0.03 and presented in Section~\ref{obs}. The result of the identification and spatial distribution of C$_{2}$H$_{5}$CN is shown in Section~\ref{res}. The discussion and conclusion of the detection of the C$_{2}$H$_{5}$CN towards G10.47+0.03 are shown in Section~\ref{dis} and Section~\ref{conclu}.

\begin{figure*}
	\centering
	\includegraphics[width=0.92\textwidth]{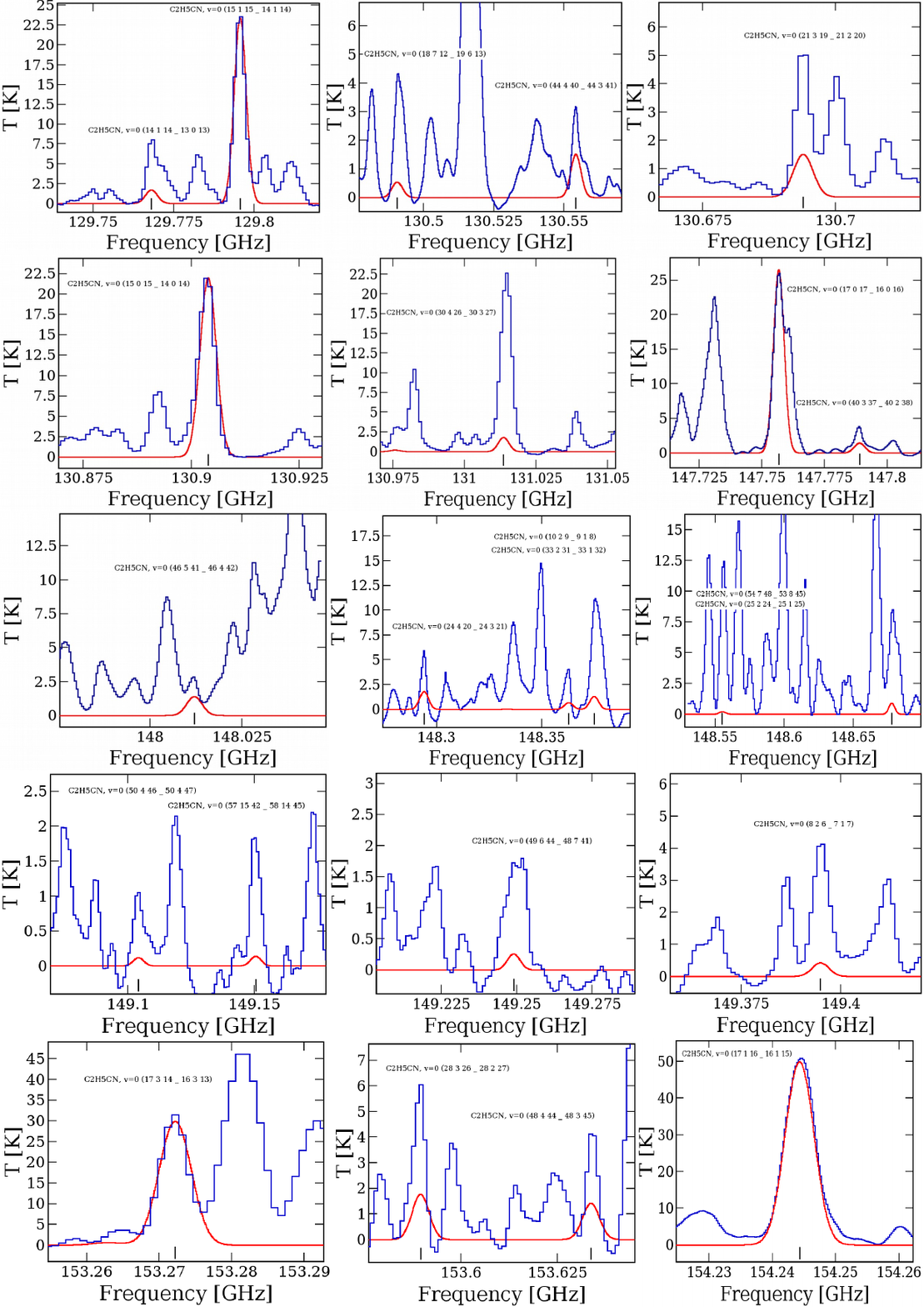}
	\caption{Rotational emission lines of C$_{2}$H$_{5}$CN with different molecular transitions towards the G10.47+0.03 in the frequency ranges of 130.234--130.949 GHz, 147.692--149.044 GHz, 153.486--154.482 GHz, and 158.942--160.155 GHz. The blue spectra indicate the observed millimeter-wavelength molecular spectra of G10.47+0.03, while the red synthetic spectra indicate the best-fitting LTE model over the observed spectra of C$_{2}$H$_{5}$CN. The black lines indicate the rest frequency positions of the C$_{2}$H$_{5}$CN transitions.}
	\label{fig:emission} 
\end{figure*}

\begin{figure*}
\text{{\large Figure~2 continued.}}
	\centering
	\includegraphics[width=0.95\textwidth]{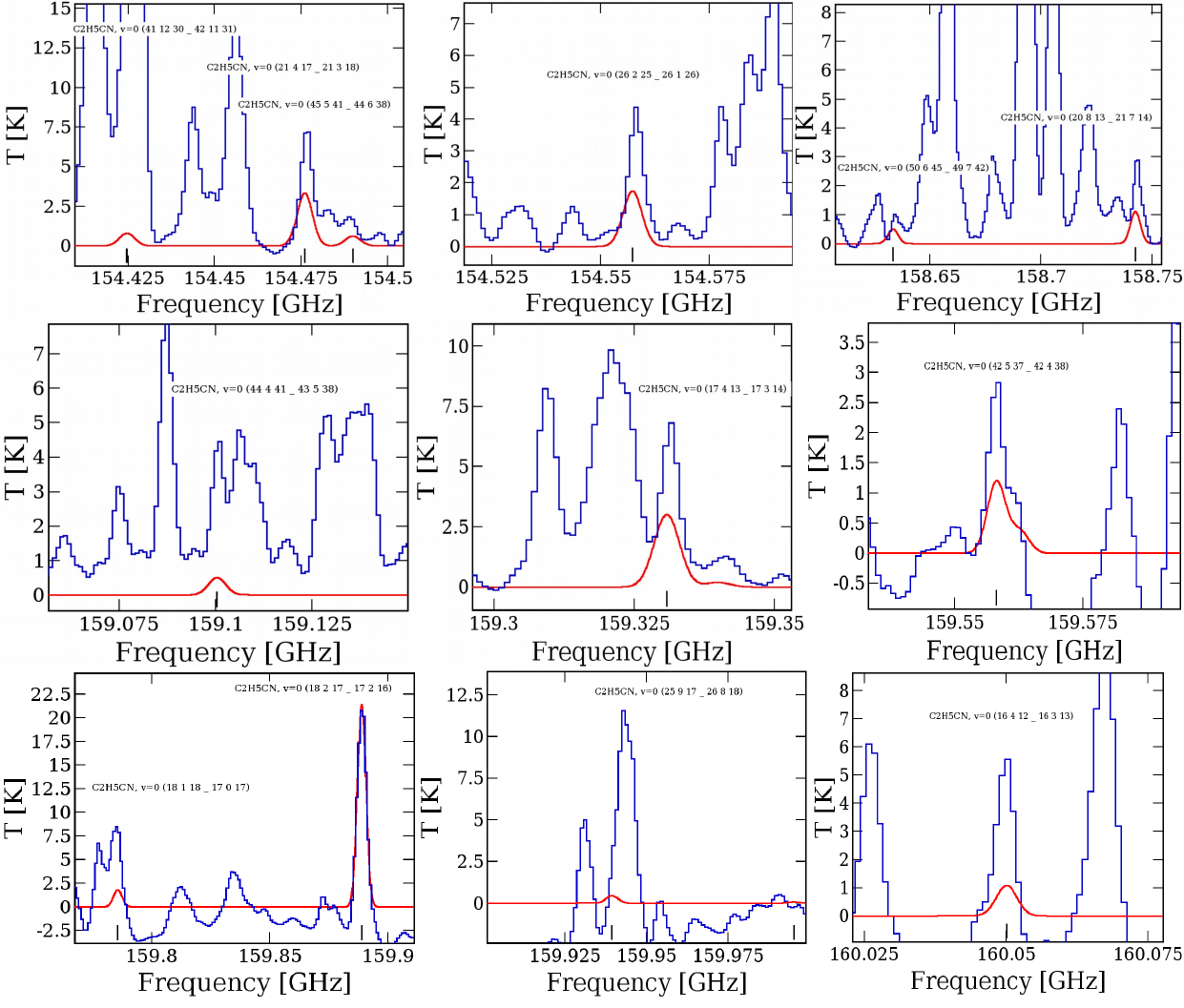}
\end{figure*}

\begin{table*}
	\centering
	\caption{Summary of the LTE fitting line properties of the C$_{2}$H$_{5}$CN  towards the G10.47+0.03}
	\begin{adjustbox}{width=0.95\textwidth}
		\begin{tabular}{ccccccccccccccccc}
			\hline 
			Observed frequency &Transition & $E_{u}$ & $A_{ij}$ &Gup&Optical depth&Remark\\
			(GHz) &(${\rm J^{'}_{K_a^{'}K_c^{'}}}$--${\rm J^{''}_{K_a^{''}K_c^{''}}}$) &(K)&(s$^{-1}$)& & ($\tau$) & \\
			\hline
			129.768&14(1,14)--13(0,13)&44.86&1.33$\times$10$^{-5}$&29&7.26$\times$10$^{-3}$&Blended with \ce{CH3OCH3}\\
			
			129.795&15(1,15)--14(1,14)&51.09&1.82$\times$10$^{-4}$&31&   1.04$\times$10$^{-1}$&Non blended\\
			
			130.490$^{*}$&18(7,12)--19(6,13)&127.98&2.17$\times$10$^{-6}$&37&  1.14$\times$10$^{-3}$&Blended with HOCO\\
			
			130.554&44(4,40)--44(3,41)&448.33&1.47$\times$10$^{-5}$&89& 6.36$\times$10$^{-3}$&Blended with \ce{DNO3}\\
			
			130.693&21(3,19)--21(2,20)&109.42&9.84$\times$10$^{-6}$&43& 6.36$\times$10$^{-3}$&Blended with \ce{NCHCCO}\\
			
			130.903&15(0,15)--14(0,14)&50.79&1.87$\times$10$^{-4}$&31& 1.05$\times$10$^{-1}$&Non blended\\
			
			131.013&30(4,26)--30(3,27)&218.54&1.30$\times$10$^{-5}$&61&  8.22$\times$10$^{-3}$&Blended with \ce{SO2}\\
			
			147.756&17(0,17)--16(0,16)&64.57&2.69$\times$10$^{-4}$&35& 1.43$\times$10$^{-1}$&Non blended\\
			
			147.788&40(3,37)--40(2,38)&367.98&1.62$\times$10$^{-5}$&81& 7.26$\times$10$^{-3}$&Blended with \ce{CH3COOH}\\
			
			148.012&46(5,41)--46(4,42)&495.98&2.13$\times$10$^{-5}$&91& 7.13$\times$10$^{-3}$&Blended with \ce{CH3CDO}\\
			
			148.294&24(4,20)--24(3,21)&147.04&1.63$\times$10$^{-5}$&49&  9.16$\times$10$^{-3}$&Blended with \ce{CH3COOH}\\
			
			148.362&10(2,9)--9(1,8)&28.05&9.75$\times$10$^{-6}$&21& 3.48$\times$10$^{-3}$&Blended with \ce{CHD2CN}\\
			
			148.374&33(2,31)--33(1,32)&249.49&1.24$\times$10$^{-5}$&67& 6.78$\times$10$^{-3}$&Blended with \ce{CH3NCO}\\
			
			148.554&54(7,48)--53(8,45)&694.07&4.77$\times$10$^{-6}$&109& 1.58$\times$10$^{-4}$&Blended with \ce{CH3OCHO}\\
			
			148.677&25(2,24)--25(1,25)&143.02&7.74$\times$10$^{-6}$&51& 4.55$\times$10$^{-3}$&Blended with \ce{H2CO}\\
			
			149.104&50(4,46)--50(4,47)&573.32&2.20$\times$10$^{-6}$&101& 6.07$\times$10$^{-4}$&Blended with \ce{CH3COOH}\\
			
			149.150$^{*}$&57(15,42)--58(14,45)&959.92&4.09$\times$10$^{-6}$&115& 3.45$\times$10$^{-4}$&Blended with c-\ce{C3H2}\\
			
			149.249&49(6,44)--48(7,41)&547.96&4.79$\times$10$^{-6}$&99& 1.31$\times$10$^{-3}$&Blended with \ce{HC5N}\\
			
			149.394&8(2,6)--7(1,7)&19.98&7.34$\times$10$^{-6}$&17& 2.15$\times$10$^{-3}$&Blended with \ce{CH3COOH}\\
			
			153.272&17(3,14)--16(3,13)&75.97&2.92$\times$10$^{-4}$&35& 1.39$\times$10$^{-1}$&Non blended\\
			
			153.389&28(3,26)--28(2,27)&184.63&1.44$\times$10$^{-5}$&57& 7.72$\times$10$^{-3}$&Blended with \ce{CH3COOH}\\
			
			153.633&48(4,44)--48(3,45)&529.95&2.14$\times$10$^{-5}$&97& 6.18$\times$10$^{-3}$&Blended with \ce{CH3COOH}\\
			
			154.244&17(1,16)--16(1,15)&68.17&3.06$\times$10$^{-4}$&35& 1.47$\times$10$^{-1}$&Non blended\\
			
			154.424&41(12,30)--42(11,31)&529.99&4.29$\times$10$^{-6}$&86& 1.05$\times$10$^{-3}$&Blended with \ce{CH3OH}\\
			
			154.476&21(4,17)--21(3,18)&117.24&1.76$\times$10$^{-5}$&43& 8.83$\times$10$^{-3}$&Blended with \ce{CHCNH}$^{+}$\\
			
			154.489&45(5,41)--44(6,38)&474.03&5.07$\times$10$^{-6}$&91& 1.63$\times$10$^{-3}$&Blended with HCOCN\\
			
			154.557&26(2,25)--26(1,26)&154.02&8.40$\times$10$^{-6}$&53&  4.59$\times$10$^{-3}$&Blended with \ce{CH3COOH}\\
			
			158.633&50(6,45)--49(7,42)&589.55&5.73$\times$10$^{-6}$&101& 1.32$\times$10$^{-3}$&Blended with \ce{CH3COOH}\\
			
			158.742$^{*}$&20(8,13)--21(7,14)&161.40&3.73$\times$10$^{-6}$&41& 2.28$\times$10$^{-3}$&Blended with \ce{HC5N}\\
			
			159.100&44(4,41)--43(5,38)&444.13&3.99$\times$10$^{-6}$&89& 1.46$\times$10$^{-3}$&Blended with \ce{C2H5OH}\\
			
			159.330&17(4,13)--17(3,14)&83.61&1.84$\times$10$^{-5}$&35& 7.89$\times$10$^{-3}$&Blended with \ce{CH3COOH}\\
			
			159.558&42(5,37)--42(4,38)&417.78&2.43$\times$10$^{-5}$&85&  2.98$\times$10$^{-3}$&Blended with c-\ce{C3HCN}\\
			
			159.785&18(1,18)--17(0,17)&72.24&2.71$\times$10$^{-5}$&37& 1.27$\times$10$^{-2}$&Blended with \ce{CH3OCHO}\\
			
			159.888&18(2,17)--17(2,16)&77.61&3.39$\times$10$^{-4}$&37& 1.55$\times$10$^{-1}$&Non blended\\
			
			159.939$^{*}$&25(9,17)--26(8,18)&229.71&4.16$\times$10$^{-6}$&51& 6.52$\times$10$^{-3}$&Blended with \ce{HC5N}\\
			
			160.050&16(4,12)--16(3,13)&76.29&1.84$\times$10$^{-5}$&33& 8.56$\times$10$^{-3}$&Blended with \ce{CH3COOH}\\
			
			\hline
		\end{tabular}	
	\end{adjustbox}
	\\
	${*}$ -- Transitions contain double with a frequency difference of less than 100 kHz. The second molecular transition of \ce{C2H5CN} is not shown.\\
	\label{tab:MOLECULAR DATA}
\end{table*}

\begin{table}{}
	\centering
	\caption{Estimated emitting regions of \ce{C2H5CN} towards the G10.47+0.03.
	}
	\begin{adjustbox}{width=0.48\textwidth}
		\begin{tabular}{cccccccccccc}
			\hline
			Frequency&Transition&$E_{up}$&Emitting region\\
			
			(GHz)&(${\rm J^{'}_{K_a^{'}K_c^{'}}}$--${\rm J^{''}_{K_a^{''}K_c^{''}}}$)&(K)&[$^{\prime\prime}$]\\
			\hline
			129.795&15(1,15)--14(1,14)&51.09&1.52 \\
			
			130.903&15(0,15)--14(0,14)&50.79&1.49 \\
			
			147.756&17(0,17)--16(0,16)&64.57&1.51\\
			
			153.272&17(3,14)--16(3,13)&75.97&1.54 \\
			
			154.244&17(1,16)--16(1,15)&68.17&1.39\\
			
			159.888&18(2,17)--17(2,16)&77.61&1.36\\
			\hline
		\end{tabular}
	\end{adjustbox}

	\label{tab:spatial}
\end{table}

\section{Observations and data reductions}
\label{obs}

We have used the cycle 4 archival data of the hot molecular core G10.47+0.03, which was observed using the high-resolution Atacama Large Millimeter/Submillimeter Array (ALMA) (\#2016.1.00929.S., PI: Ohishi, Masatoshi). The G10.47+0.03 was located at the observed phase center of ($\alpha,\delta$)$_{\rm J2000}$ = 18:08:38.232, --19:51:50.400. The observation of G10.47+0.03 was carried out on 28 January 2017, 5 March 2017, 6 March 2017, and 7 March 2017, using the thirty-nine, forty, forty-one, and thirty-nine antennas, respectively. During the observations, the minimum baseline of the antennas was 15 m, and the maximum baseline of the antennas was 331 m. The observations were made with ALMA-band 4 receivers with spectral ranges of 129.50 GHz--160.43 GHz and a corresponding angular resolution of 1.67$^{\prime\prime}$--1.76$^{\prime\prime}$. The corresponding spectral resolutions of the observed data were 1128.91 kHz and 488.28 kHz. The flux calibrator was taken as J1733--1304, the bandpass calibrator was taken as J1924--2914, and the phase calibrator was taken as J1832--2039. The observation details are shown in Table~\ref{tab:data}.

We use the Common Astronomy Software Application ({\tt CASA 5.4.1}) for the reduction of the ALMA data \citep{m07}. We use the Perley-Butler 2017 flux calibrator model to scale the continuum flux density of the flux calibrator using the CASA task {\tt SETJY} with 5\% accuracy \citep{pal17}. We use the task {\tt MSTRANSFORM} to split the target data with all rest frequencies after the initial data reduction. After the splitting of the target data, we create the continuum images of G10.47+0.03 in all rest frequencies using CASA task {\tt TCLEAN} with {\tt hogbom} deconvolar. Now we perform the continuum subtraction procedure from the UV plane using the task {\tt UVCONTSUB}. After the continuum subtraction, we use the {\tt TCLEAN} task with a Briggs weighting robust parameter of 0.5 to make the spectral data cubes of G10.47+0.03. Finally, we apply the {\tt IMPBCOR} task in CASA for the primary beam pattern correction in the synthesised image. Our data analysis procedure is similar to \citet{man22c} and \citet{man22d}.

\begin{figure*}
	\centering
	\includegraphics[width=0.95\textwidth]{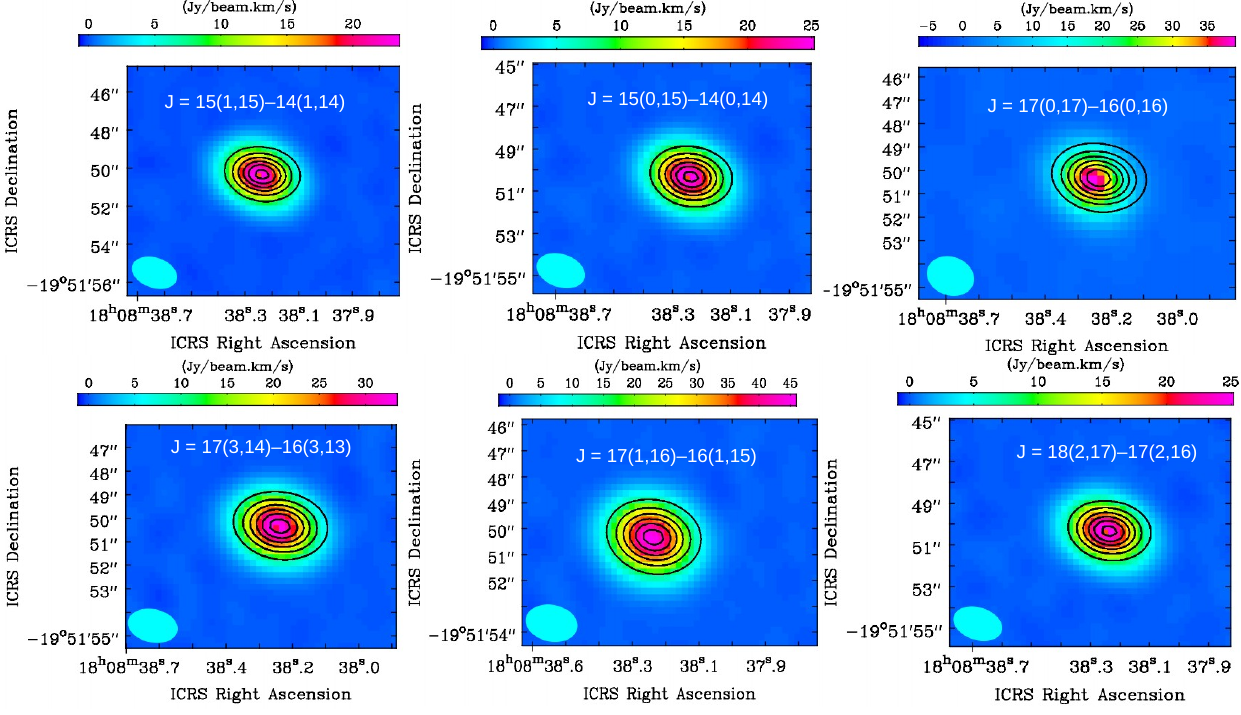}
	\caption{ Integrated emission maps of non-blended transitions of C$_{2}$H$_{5}$CN towards the G10.47+0.03. The integrated emission maps of C$_{2}$H$_{5}$CN are overlaid with the 130.773 GHz (2.29 mm) continuum emission map (black contour). The contour levels start at 3$\sigma$, where $\sigma$ is the RMS of each emission map, and the contour levels increase by a factor of $\surd$2. The cyan circle indicate the synthesised beam of the C$_{2}$H$_{5}$CN integrated emission maps.}
	\label{fig:map}
\end{figure*}

\begin{table*}
	\centering
	\caption{Summary of the Gaussian fitting line properties of the non-blended emission lines of C$_{2}$H$_{5}$CN  towards the G10.47+0.03}
	\begin{adjustbox}{width=0.9\textwidth}
		\begin{tabular}{ccccccccccccccccc}
			\hline 
			Observed frequency &Transition & $E_{u}$ & $A_{ij}$ &FWHM &$\rm{\int T_{mb}dV}$ \\
			
			(GHz) &(${\rm J^{'}_{K_a^{'}K_c^{'}}}$--${\rm J^{''}_{K_a^{''}K_c^{''}}}$) &(s$^{-1}$) &(K)& (km s$^{-1}$) &(K km s$^{-1}$) & \\
			\hline
			129.795&15(1,15)--14(1,14)&51.09&1.82$\times$10$^{-4}$&10.02$\pm$0.69&222.05$\pm$10.24\\
			
			130.903&15(0,15)--14(0,14)&50.79&1.87$\times$10$^{-4}$&10.01$\pm$0.75&193.03$\pm$12.73\\
			
			147.756&17(0,17)--16(0,16)&64.57&2.69$\times$10$^{-4}$&10.81$\pm$0.55&212.83$\pm$22.62\\
			
			153.272&17(3,14)--16(3,13)&75.97&2.92$\times$10$^{-4}$&9.82$\pm$0.87&279.48$\pm$39.65\\
			
			154.244&17(1,16)--16(1,15)&68.17&3.06$\times$10$^{-4}$&10.41$\pm$0.96&342.82$\pm$29.96\\
			
			159.888&18(2,17)--17(2,16)&77.61&3.39$\times$10$^{-4}$&10.29$\pm$0.67&206.34$\pm$35.74\\
			\hline
		\end{tabular}	
	\end{adjustbox}
	\label{tab:gaussianfittedre}
\end{table*}

\section{Result}
\label{res}
\subsection{Millimeter wavelength continuum emission towards the G10.47+0.03}
We presented the millimeter wavelength continuum emission maps towards the G10.47+0.03 at frequencies of 130.234 GHz, 130.320 GHz, 130.773 GHz, 130.949 GHz, 147.692 GHz, 147.992 GHz, 148.237 GHz, 149.044 GHz, 153.486 GHz, 153.597 GHz, 154.038 GHz, 154.482 GHz, 158.942 GHz, 159.414 GHz, 159.928 GHz, and 160.155 GHz in Figure~\ref{fig:continuum}, where the surface brightness colour scale has the unit of Jy beam$^{-1}$. We estimate the integrated flux density, peak flux density, synthesised beam size, position angle, and RMS of the G10.47+0.03 to fit a 2D Gaussian over the continuum images using CASA task {\tt IMFIT}. The estimated continuum image properties of the G10.47+0.03 are shown in Table.~\ref{tab:cont}. After fitting the 2D Gaussian, we observe that the synthesised beam size of the continuum image of G10.47+0.03 is not sufficient to resolve the continua. Previously, \citet{rol11} detected the submillimeter continuum emission from the G10.47+0.03 in the frequency range of 201--691 GHz with the variation of flux density as 6--95 Jy, corresponding to the spectral index 2.8.

\begin{figure*}
	\centering
	\includegraphics[width=0.8\textwidth]{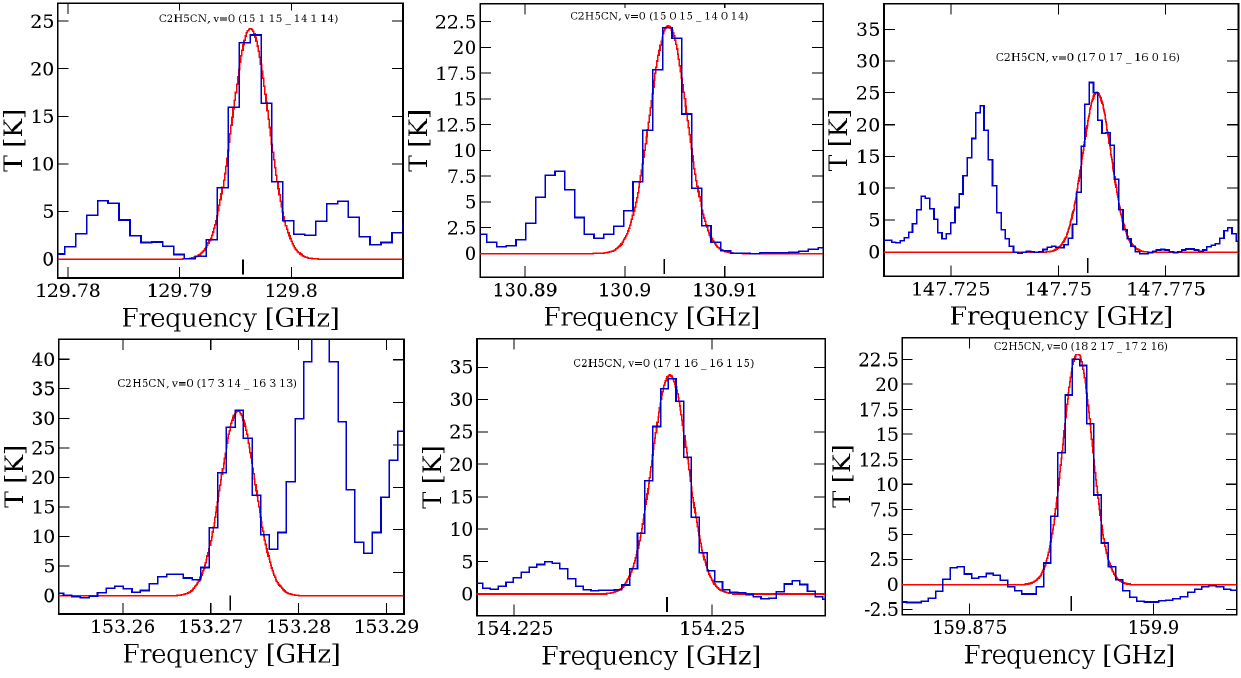}
	\caption{Non-blended rotational emission lines of C$_{2}$H$_{5}$CN with different molecular transitions towards the G10.47+0.03. The blue spectra indicate the observed millimeter-wavelength molecular spectra of G10.47+0.03, while the red synthetic spectra indicate the best-fitting Gaussian model, which was fitted over the non-blended observed spectra of C$_{2}$H$_{5}$CN. The black lines indicate the rest frequency positions of the C$_{2}$H$_{5}$CN transitions.}
	\label{fig:gaussian} 
\end{figure*}

\subsection{Identification of C$_{2}$H$_{5}$CN towards the G10.47+003}
We generate the millimeter-wavelength chemically rich spectra from the continuum-subtracted spectral data cubes of G10.47+0.03 to create a circular region with a diameter of 2.5$^{\prime\prime}$ centred at RA (J2000) = 18$^{h}$08$^{m}$38$^{s}$.232, Dec (J2000) = --19$^\circ$51$^{\prime}$50$^{\prime\prime}$.440. The systematic velocity ($V_{LSR}$) of the G10.47+0.03 is 68.50 km s$^{-1}$ \citep{rol11}. For identification of the rotational emission lines of \ce{C2H5CN} towards the G10.47+0.03, we use the local thermodynamic equilibrium (LTE) model with the Cologne Database for Molecular Spectroscopy (CDMS) \citep{mu05} spectroscopic molecular database. We have used the LTE-RADEX module in CASSIS for LTE computing \citep{vas15}. The LTE assumption is valid in the warm inner region of G10.47+0.03 because the gas density of the hot core region of G10.47+0.03 is 7$\times$10$^{7}$ cm$^{-3}$ \citep{rol11}. After the LTE analysis, we detect a total of thirty-six transition lines of \ce{C2H5CN} within the observable frequency ranges. There are no missing transitions of \ce{C2H5CN} between the frequency ranges of 130.234--130.949 GHz, 147.692--149.044 GHz, 153.486--154.482 GHz, and 158.942--160.155 GHz. At the full-beam offset position, the best-fit column density of \ce{C2H5CN} is 8.02$\times$10$^{16}$ cm$^{-2}$ with excitation temperature 300 K and source size 2.5$^{\prime\prime}$ (beam size of the data cubes). The FWHM of the LTE model spectra of \ce{C2H5CN} is 10.2 km s$^{-1}$. The LTE-fitted rotational emission lines of \ce{C2H5CN} are shown in Figure~\ref{fig:emission}. 

Recently, \citet{mondal23} also analysed the molecular spectra of G10.47+0.03 using the ALMA, and they estimated the column density of \ce{C2H5CN} to be (1.7$\pm$0.10)$\times$10$^{17}$ cm$^{-2}$ with an excitation temperature of 150 K. \citet{mondal23} fit above the second-order polynomial to reduce the noise level of the spectra, which contributed to the measurement of a higher column density of \ce{C2H5CN}. \citet{mondal23} could not detect all the transition lines of \ce{C2H5CN} towards the G10.47+0.03 within the observable frequency ranges. \citet{mondal23} determined the temperature of \ce{C2H5CN} using both LTE spectra and a rotational diagram model. There is a discrepancy in the temperature measurement of the \ce{C2H5CN} emitting region by the two methods used in \citet{mondal23}. Using LTE modelling, \citet{mondal23} estimated the temperature of \ce{C2H5CN} to be 150 K, which indicates that the molecule is arising from the warm-inner region of the hot core because the temperature of the hot core region of G10.47+0.03 is above 100 K. But, using the rotational diagram method, \citet{mondal23} found the temperature of the \ce{C2H5CN} emitting region to be 92 K, indicating that the molecule is arising from the cold region of the hot core. So, the estimated temperature of \ce{C2H5CN} by \citet{mondal23} is confusing. Earlier, \citet{rol11} showed that the temperature of \ce{C2H5CN} emiting region is above 200 K.

After the identification of the rotational emission lines of \ce{C2H5CN} from the millimeter wavelength spectra of G10.47+0.03, we obtain the molecular transitions ({${\rm J^{'}_{K_a^{'}K_c^{'}}}$--${\rm J^{''}_{K_a^{''}K_c^{''}}}$}), upper state energy ($E_u$) in K, Einstein coefficients ($A_{ij}$) in s$^{-1}$, $G_{up}$, and optical depth ($\tau$). After the LTE analysis, we observe  that J = 15(1,15)--14(1,14), J = 15(0,15)--14(0,14), J = 17(0,17)--16(0,16), J = 17(3,14)--16(3,13), J = 17(1,16)--16(1,15), and J = 18(2,17)--17(2,16) transition lines of \ce{C2H5CN} are non-blended. We also observe that all non-blended emission lines of \ce{C2H5CN} are properly fitted with the LTE model, but the blended lines are not properly fitted with the LTE model. The summary of the LTE fitted line properties of C$_{2}$H$_{5}$CN is shown in Table~\ref{tab:MOLECULAR DATA}.

\subsection{Spatial distribution of \ce{C2H5CN}}
We create the integrated emission maps of non-blended transitions of C$_{2}$H$_{5}$CN towards the G10.47+0.03. To create the integrated emission maps, we use the task {\tt IMMOMENTS} in CASA. During the run of the task {\tt IMMOMENTS}, we define the channel ranges of the data cubes where the emission lines of C$_{2}$H$_{5}$CN are detected. The integrated emission maps of C$_{2}$H$_{5}$CN are shown in Figure~\ref{fig:map}, which are overlaid on the 2.29 mm continuum emission map of G10.47+0.03. After overlaying the continuum emission map over the integrated emission maps of C$_{2}$H$_{5}$CN, we found that the emission maps of C$_{2}$H$_{5}$CN have a peak at the position of the continuum. The resultant non-blended integrated emission maps indicate that the different transitions of the C$_{2}$H$_{5}$CN molecule arise from the highly dense warm inner hot core region of G10.47+0.03. To estimate the emitting regions of C$_{2}$H$_{5}$CN, we apply the task {\tt IMFIT} for fitting the 2D Gaussian over the integrated emission maps of C$_{2}$H$_{5}$CN. The deconvolved synthesised beam size of the C$_{2}$H$_{5}$CN emitting regions is estimated by the following equation:

\begin{equation}		
\theta_{S}=\sqrt{\theta^2_{50}-\theta^2_{\text{beam}}}		
\end{equation}
where $\theta_{\text{beam}}$ is the half-power width of the synthesised beam of the C$_{2}$H$_{5}$CN integrated emission maps and $\theta_{50} = 2\sqrt{A/\pi}$ is the diameter of the circle whose area ($A$) is surrounding the $50\%$ line peak of C$_{2}$H$_{5}$CN \citep{man22c, man22d}. The estimated emitting regions of C$_{2}$H$_{5}$CN are shown in Table~\ref{tab:spatial}. The emitting regions of C$_{2}$H$_{5}$CN are observed in the range of 1.36$^{\prime\prime}$--1.54$^{\prime\prime}$. We notice that the estimated emitting regions of C$_{2}$H$_{5}$CN are comparable to or slightly greater than the synthesised beam size of the integrated emission maps, which indicates the observed transitions of C$_{2}$H$_{5}$CN are not well spatially resolved or, at best, marginally resolved. As a result, no conclusions can be drawn about the morphology of the spatial distribution of \ce{C2H5CN} towards the G10.47+0.03.

\subsection{Rotational diagram analysis of \ce{C2H5CN}}
\label{rotd}
\begin{figure*}
	\includegraphics[width=0.48\textwidth]{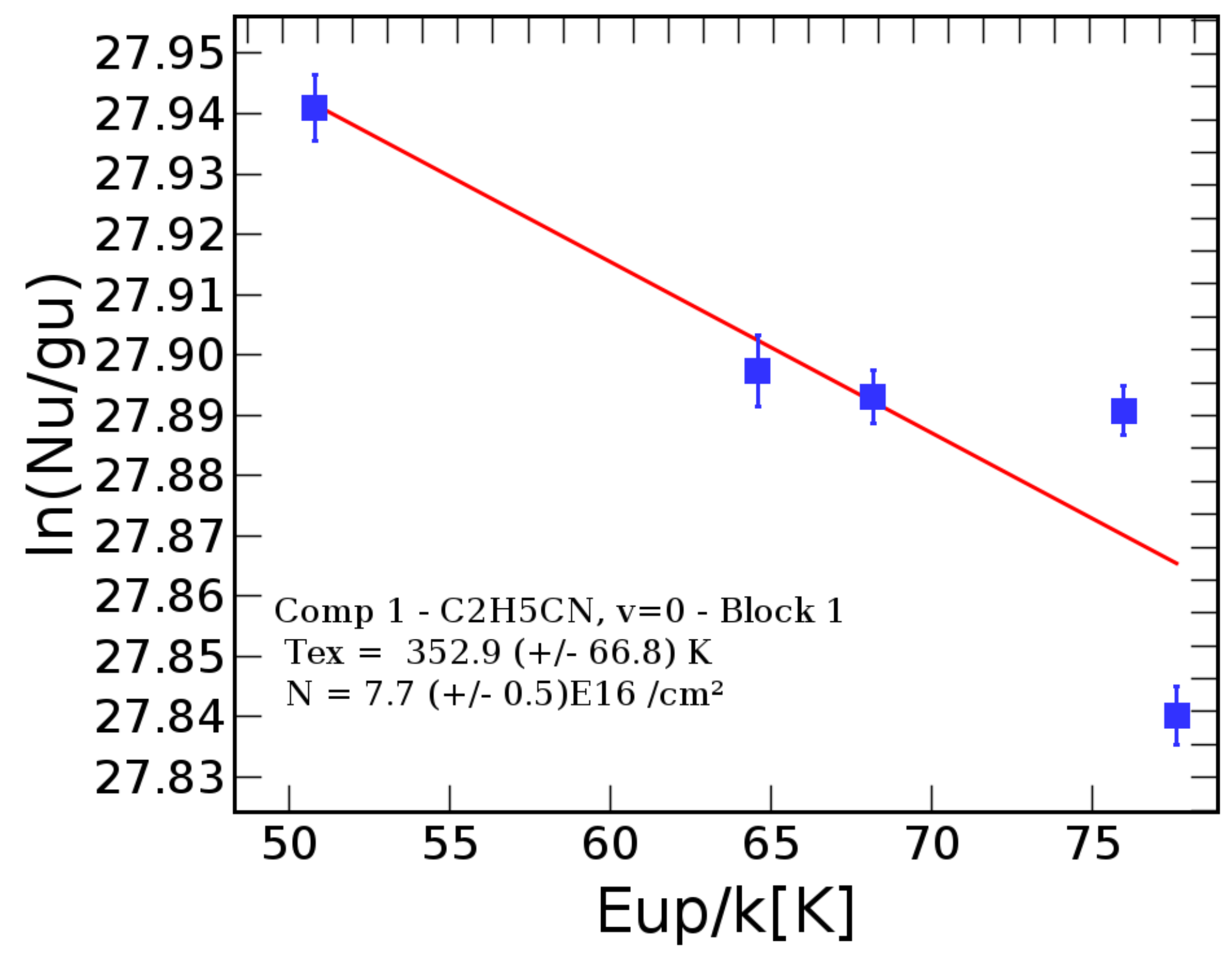}
	\centering
	\caption{Rotational diagram of \ce{C2H5CN} towards the G10.47+0.03. The blue-filled squares indicate the optically thin approximation data points and blue vertical lines represent the error bars. The best-fit column density and rotational temperature are mentioned inside the image.}
	\label{fig:rotd} 
\end{figure*}

We use the rotational diagram model for the estimation of the column density and rotational temperature of the detected non-blended rotational emission lines of \ce{C2H5CN}. Initially, we consider that the detected molecular spectra of \ce{C2H5CN} are optically thin and obey the local thermodynamic equilibrium (LTE) conditions. The LTE approximation is appropriate in the G10.47+0.03 environment because the gas density of the warm inner region of G10.47+0.03 is 7$\times$10$^{7}$ cm$^{-3}$ \citep{rol11}. The column density of the optically thin molecular emission lines can be written as \citep{gold99},

\begin{equation}
{N_u^{thin}}=\frac{3{g_u}k_B\int{T_{mb}dV}}{8\pi^{3}\nu S\mu^{2}}
\end{equation}
where $g_u$ indicates the degeneracy of the upper energy ($E_u$), $k_B$ represents the Boltzmann constant, $\rm{\int T_{mb}dV}$ indicates the integrated intensity, $\mu$ is the electric dipole moment, $S$ denote the strength of the detected molecular emission lines, and $\nu$ is the rest frequency of the identified emission lines of \ce{C2H5CN}. The equation for total column density under LTE conditions is as follows,

\begin{equation}
\frac{N_u^{thin}}{g_u} = \frac{N_{\text{total}}}{Q(T_{\text{rot}})}\exp(-E_u/k_BT_{\text{rot}})
\end{equation}
where ${Q(T_{\text{rot}})}$ denotes the partition function as a function of rotational temperature ($T_{\text{rot}}$) and $E_{u}$ denote the identified molecule's upper energy. The partition functions of \ce{C2H5CN} at 75 K, 150 K, and 300 K are 4667.9361, 13209.5867, and 37424.5763, respectively. Equation 3 can be rewritten as,

\begin{equation}
ln\left(\frac{N_u^{thin}}{g_u}\right) = ln\left(\frac{N}{Q}\right)-\left(\frac{E_u}{k_BT_{rot}}\right)
\end{equation}
Equation 4 indicates a linear connection between the values of $\ln(N_{u}/g_{u}$) and E$_{u}$ of the detected emission lines of \ce{C2H5CN}. The values $\ln(N_{u}/g_{u}$) are calculated from equation 2. The column density and rotational temperature of \ce{C2H5CN} can be estimated by fitting a straight line over the values of $ln$(N$_{u}$/g$_{u}$), which are plotted as a function of E$_{u}$. After fitting the straight line, the rotational temperature is estimated from the inverse of the slope, and the column density is estimated from the intercept of the slope. For the rotational diagram, we extract the spectral line parameters of the six non-blended rotational emission lines of \ce{C2H5CN} using the Gaussian model. We have used the Levenberg-Marquardt (LM)\footnote{\url{http://cassis.irap.omp.eu/docs/help_cassis_fit_intro.pdf}} algorithm in CASSIS for fitting the Gaussian model over the detected emission lines of \ce{C2H5CN}. We observe that the $E_{up}$ of the J = 15(1,15)--14(1,14) and J = 15(0,15)--14(0,14) transition lines of \ce{C2H5CN} are nearly similar. So, we use only the J = 15(0,15)--14(0,14) transition line for the rotational diagram because the line intensity of J = 15(0,15)--14(0,14) is higher than J = 15(1,15)--14(1,14). The best fit non-blended rotational emission lines of \ce{C2H5CN} with the Gaussian model is shown in Figure~\ref{fig:gaussian} and spectral line fitting parameters are shown in Table~\ref{tab:gaussianfittedre}. To draw the rotational diagram of non-blended transitions of \ce{C2H5CN}, we use the ROTATIONAL DIAGRAM module in CASSIS. The resultant rotational diagram is shown in Figure~\ref{fig:rotd}. In the rotational diagram, the blue error bars represent the error bar of $\ln(N_{u}/g_{u}$), which is calculated from the estimated error of $\rm{\int T_{mb}dV}$ by fitting a Gaussian model over the observed emission lines of \ce{C2H5CN}. The estimated total column density of \ce{C2H5CN} towards the G10.47+0.03 is (7.7$\pm$0.5)$\times$10$^{16}$ cm$^{-2}$ with the high rotational temperature of 352.9$\pm$66.8 K. Our estimated rotational temperature indicate that the emission lines of \ce{C2H5CN} originated from the hot core region of G10.47+0.03 because \citet{rol11} claimed the temperature of the hot core of G10.47+0.03 is above 100 K. Our estimated total column density and rotational temperature of \ce{C2H5CN} are nearly similar to the LTE-fitted column density and excitation temperature of \ce{C2H5CN} towards the G10.47+0.03. The fractional abundance of \ce{C2H5CN} with respect to {H$_{2}$} is calculated to be 5.70$\times$10$^{-9}$, and the hydrogen column density towards G10.47+0.03 is calculated to be 1.35$\times$10$^{25}$ cm$^{-2}$ \citep{gor20}.

\section{Discussion}
\label{dis}
\subsection{Comparison with observation and modelled abundance of C$_{2}$H$_{5}$CN}
We compare the estimated abundance of \ce{C2H5CN} with the three-phase warm-up chemical modelling results of \citet{gar13}, which is applied in the environment of the hot molecular cores. During the chemical modelling, \citet{gar13} assumed there would be an isothermal collapse phase after a static warm-up phase. In the first phase, the gas density increase from 3$\times$10$^{3}$ to 10$^{7}$ cm$^{-3}$ under the free-fall collapse, and the dust temperature decreases from 16 K to 8 K. In the second phase, the gas density remains constant at $\sim$10$^{7}$ cm$^{-3}$ but the dust temperature fluctuates from 8 K to 400 K. The gas density ($n_{H}$) of G10.47+0.03 is 7$\times$10$^{7}$ cm$^{-3}$ \citep{rol11} and the temperature of the warm region is $\sim$150 K \citep{rol11}. So, the three-phase warm-up chemical modelling of \cite{gar13}, which is applied towards hot molecular cores, is suitable for understanding the chemical evolution of \ce{C2H5CN} towards G10.47+0.03. In the three-phase warm-up chemical modelling, \citet{gar13} used the fast (5$\times$10$^{4}$--7.12$\times$10$^{4}$ years), medium (2$\times$10$^{5}$--2.85$\times$10$^{5}$ years), and slow (1$\times$10$^{6}$--1.43$\times$10$^{6}$ years) warm-up models based on the time scales. \cite{gar13} estimated the abundance of \ce{C2H5CN} based on a three-phase warm-up model.  After the chemical modelling, \cite{gar13} estimated that the abundance of \ce{C2H5CN} was 5.7$\times$10$^{-9}$ for the fast warm-up model, 7.7$\times$10$^{-8}$ for the medium warm-up model, and 3.0$\times$10$^{-8}$ for the slow warm-up model. We estimate that the abundance of \ce{C2H5CN} towards G10.47+0.03 is 5.70$\times$10$^{-9}$, which is similar to the simulated abundance of \ce{C2H5CN} in the fast warm-up model. Earlier, \citet{ohi19}, \citet{man22c}, and \cite{man22d} also claimed that the three-phase warm-up chemical modelling of \citet{gar13} is sufficient to understand the chemical environment of G10.47+0.03.

\subsection{Possible formation mechanism of C$_{2}$H$_{5}$CN in the hot molecular cores}
The complex nitrogen-bearing molecule \ce{C2H5CN} is produced on the grain surface of the hot molecular cores \citep{meh04, gar13, gar17, gar22}. In the first stage (the free-fall collapse phase), the peak abundance of \ce{C2H5CN} is $\sim$10$^{-9}$ \citep{gar13}. In the free-fall collapse stage, the addition of radical \ce{CH2} with the radical \ce{CH2CN} produces the radical \ce{CH2CH2CN}, and again hydrogenation of the radical \ce{CH2CH2CN} produces the low abundance of \ce{C2H5CN} ($\sim$10$^{-9}$) on the grain surface of hot molecular cores \citep{gar13}. The chemical reactions are as follows:\\\\
	\ce{CH2}+\ce{CH2CN}$\rightarrow$\ce{CH2CH2CN}~~~~~~~~~~~~(1) \\\\
	\ce{CH2CH2CN}+H$\rightarrow$\ce{C2H5CN}~~~~~~~~~~~~~~(2) \\\\
Reaction 1 indicates that the addition of radical \ce{CH2} and radical \ce{CH2CN} (alternatively, radical-radical reactions) is barrierless and exothermic \citep{sin21}. Our estimated abundance of \ce{C2H5CN} towards the G10.47+0.03 is 5.70$\times$10$^{-9}$, which indicates that reactions 1 and 2 are responsible for the production of the \ce{C2H5CN} in the grain surface of the G10.47+0.03. \cite{gar17} and \cite{gar22} also used this reaction to simulate the abundance of \ce{C2H5CN} in the environment of the other hot molecular cores in the free-fall collapse and warm-up phases.

\section{Conclusion}
\label{conclu}
We analyse the ALMA band 4 data of the hot molecular core G10.47+0.03 and extract the millimeter wavelength rotational molecular lines. The main conclusions of this work are as follows: \\\\
1. We detect a total of thirty-six rotational emission lines of \ce{C2H5CN} towards the G10.47+0.03 using the ALMA band 4 observation. \\\\
2. The estimated column density of \ce{C2H5CN} towards the G10.47+0.03 is (7.7$\pm$0.5)$\times$10$^{16}$ cm$^{-2}$ with a rotational temperature of 352.9$\pm$66.8 K. The derived abundance of \ce{C2H5CN} with respect to \ce{H2} towards G10.47+0.03 is 5.70$\times$10$^{-9}$. \\\\
3. We create the integrated emission maps of non-blended transitions of \ce{C2H5CN}. From the emission maps, we observe that the non-blended transitions of \ce{C2H5CN} is arising from the warm-inner region of the G10.47+0.03.\\\\
4. We also compare our estimated abundance of \ce{C2H5CN} with the existing three-phase warm-up chemical modelling abundance of \ce{C2H5CN}, which is applied towards particularly hot molecular cores. After the comparison, we found that the derived abundance of \ce{C2H5CN} is nearly similar to the modelled abundance of \ce{C2H5CN} under the fast warm-up conditions. \\\\
5. We also discuss the possible formation mechanism of \ce{C2H5CN} towards the G10.47+0.03 and we claim the barrierless and exothermic radical-radical reaction between \ce{CH2} and \ce{CH2CN} is responsible for the production of the \ce{C2H5CN} in the grain surface of G10.47+0.03. The identification of the emission lines of \ce{C2H5CN} towards the G10.47+0.03 indicates that more complex nitrogen-bearing molecules like propyl cyanide (\ce{C3H7CN}) are  also detectable towards the G10.47+0.03 using the ALMA.

\section*{Acknowledgments}{We thank the anonymous referee for the helpful comments that improved the manuscript.
A.M. acknowledges the Swami Vivekananda Merit-cum-Means Scholarship (SVMCM), Government of West Bengal, India, for financial support for this research. This paper makes use of the following ALMA data: ADS /JAO.ALMA\#2016.1.00929.S. ALMA is a partnership of ESO (representing its member states), NSF (USA), and NINS (Japan), together with NRC (Canada), MOST and ASIAA (Taiwan), and KASI (Republic of Korea), in co-operation with the Republic of Chile. The Joint ALMA Observatory is operated by ESO, AUI/NRAO, and NAOJ. }\\\\

\section*{Data availability}{The data that support the plots within this paper and other findings of this study are available from the corresponding author upon reasonable request. The raw ALMA data are publicly available at \url{https://almascience.nao.ac.jp/asax/} (project id:  2016.1.00929.S).} 

\section*{Funding} No funds or grants were received during the preparation of this manuscript.

\section*{Conflicts of interest}
The authors declare no conflict of interest.

\section*{Author Contributions}
S.P. conceptualize the project. A.M. analysed the ALMA data and identify the emission lines of \ce{C2H5CN} from the G10.47+0.03. A.M analyses the rotational diagram to derive the column density and rotational temperature of \ce{C2H5CN}. A.M. and S.P. wrote the main manuscript text. All authors reviewed the manuscript.

\makeatletter
\let\clear@thebibliography@page=\relax
\makeatother

\end{document}